# Combined Channel and Spatial Attention-based Stereo Endoscopic Image Super-Resolution

Mansoor Hayat
*Department of Electrical Engineering*
*Chulalongkorn University*
Bangkok, Thailand
6471015721@student.chula.ac.th

Supavadee Armvith
*Multimedia Data Analytics and*
*Processing Research Unit*
*Department of Electrical Engineering*
*Chulalongkorn University*
Bangkok, Thailand
supavadee.a@chula.ac.th

Dr. Titipat Achakulvisut
*Department of Biomedical Engineering*
*Mahidol University*
Bangkok, Thailand
titipat.ach@mahidol.edu

*Abstract*— Stereo Imaging technology integration into medical diagnostics and surgeries brings a great revolution in the field of medical sciences. Now, surgeons and physicians have better insight into the anatomy of patients' organs. Like other technologies, stereo cameras have limitations, e.g., low resolution (LR) and blurry output images. Currently, most of the proposed techniques for super-resolution focus on developing complex blocks and complicated loss functions, which cause high system complexity. We proposed a combined channel and spatial attention block to extract features incorporated with a specific but very strong parallax attention module (PAM) for endoscopic image super-resolution. The proposed model is trained using the da Vinci dataset on scales 2 and 4. Our proposed model has improved PSNR up to 2.12 dB for scale 2 and 1.29 dB for scale 4, while SSIM is improved by 0.03 for scale 2 and 0.0008 for scale 4. By incorporating this method, diagnosis and treatment for endoscopic images can be more accurate and effective.

## I. INTRODUCTION

Due to various circumstances, including the medical equipment and external environment, medical imaging procedures frequently produce damaged images. Medical Image Super-Resolution is a crucial area of research, as high-resolution (HR) medical images play a vital role in diagnosing and treating various medical conditions. Using HR medical images improves the accuracy of diagnosis and treatment planning, leading to better patient outcomes.

The utility of a medical image is dependent on its quality. Extended scans leave images prone to motion artifacts. The SR problem is challenging due to the diversity of solutions for the inverse operation for a given pixel. More than one solution algorithm can be applied to the SR problem.

Modern and advanced imaging technologies, now conventional surgeries, are also improved with camera-based real-time imaging, which provides accurate visualization of patients' organs for diagnostics and surgeries. This makes the treatment more accurate with precise conditions [1]. To visualize the position and condition of diseased organs in Cardiology, Gynecology, Transplant, Gastrointestinal, Orthopedic, Colon, Rectal Surgery, and others, utilize medical imagery. Monoscopic images are usually the output of most medical image cameras [2]. Left and right LR image pairs can help construct HR images, known as stereo image super-resolution. Intra-view and cross-view information are leveraged to achieve stereo image super-resolution. This results in achieving high-quality SR images [3].

The field of view for the monocular image is a limit of the monoscopic and still relies on passive synthesis related to depth-of-field information of human eyes. So, it lacks in providing complete three-dimensional data for robotic surgery. This limitation has caused implications for the intelligent direction of surgical robot development because it restricts the ability of robots to accurately perceive and respond to the surgical environment [4-6]. Stereo vision, which utilizes two cameras to capture deep information, has become very popular in mobile phones and autonomous vehicles. The proposed method can enhance the image's resolution using pairs [7]. The endoscope in medical sciences is highly effective for diagnosing diseases within internal organs [8]. The medical endoscope achieved widespread use in the patients' examination, disease diagnosis, and disease treatment within the alimentary system, including the esophagus [9], stomach [10], and intestines [11]. This can accurately investigate pathological changes within internal organs without causing radiation damage.

Three main classes of super-resolution methods: interpolation-based [12], reconstruction-based [13], and learning-based approaches [14]. Interpolation-based methods are commonly utilized for creating zoom-in images due to their simplicity. They use adaptive structure kernels, local covariance coefficients, or fixed function kernels. However, these methods are more likely to result in visual abnormalities such as blurring, blocking, aliasing, and jagged image artifacts as the magnification rises. MSISRD [15] proposed a novel super-resolution (SR) technique using deep learning. Unlike existing methods, it utilizes a multi-scale inception-based network and avoids interpolation techniques that introduce noise. This method employs deconvolution and asymmetric convolution blocks for accurate and efficient reconstruction.

Reconstruction-based techniques consider LR images as the consequence of various degradation factors, such as downsampling and blurring. They strongly emphasize the reconstruction constraint, ensuring that the goal HR image after downsampling corresponds to its LR counterpart. However, reconstructed-based approaches frequently produce HR with ringing effects and edges that are overly sharp and unnatural, particularly along crucial edges.

Learning-based super-resolution techniques have provided substantial progress in the last couple of years. Using an image database or the actual image, this method determines the connection between HR and LR images and produces the HR image as an a priori constraint. The outstanding performance of learning-based SR techniques has drawn a lot of interest. Deep learning, sparse coding, and neighbor embedding (NE) are the three primary categories of learning-based algorithms.

Overall, the endoscopic view helps to reconstruct the 3D structure of the organ under observation. This is not only helpful in treatment and diagnostics but also very useful for students related to medical sciences to understand the anatomy, structure, abnormalities, and different situations associated with human organs.

We have proposed a feature extraction block that can tweak the efficiency of the overall model. The proposed block is composed of channel and spatial attention blocks. The channel attention block is used to learn the importance of each channel, while the spatial attention block is used to learn the importance of each spatial location in each image. The combination of channel and spatial attention blocks allows the proposed feature extraction block to extract features more effectively than traditional ones. This feature extraction block is then integrated with PAM for feature interaction and finally upsampling is applied to extract super-resolute images. We have called this model Combined Channel and Spatial Attention-based Stereo Endoscopic Image Super-Resolution (CCSBESR).

## II. RELATED WORK

Super-resolution methods are commonly categorized as Single Image Super-Resolution (SISR) [12], Multi-Image Super-Resolution (MISR) [13], Medical Image Super-Resolution [16], and Video Super-resolution (VSR) [14]. While SISR is useful for single independent images, it fails to use the continuity between numerous images in video data, such as endoscopic videos, resulting in suboptimal super-resolution results. MISR incorporates multiple LR images to produce a single HR image.

These methods take pairs of LR and HR patches in pairs and attempt to learn a mapping to translate the LR patches into HR ones. Example pair techniques can be created for both general images and a specific type of image, say medical images, depending on the training set of examples provided. The state-of-the-art representative example-based super-resolution model is sparse coding.

Researchers have developed several fresh methods for deep convolutional neural networks for super-resolution reconstruction to overcome these issues. Hu et al. [17] altered the network architecture to simplify performance and training. To increase the correlation between neighboring feature information and boost the overall quality of the reconstructed image, another method includes adding context information to the network. Although there have been breakthroughs in the field, more can be done to enhance it. Therefore research is constantly being done to develop methods for super-resolution reconstruction.

### A. Single Image Super-Resolution (SISR)

The SISR is a crucial problem that researchers have been investigating for several decades. Recently, SISR approaches using deep learning demonstrated promising outcomes in terms of reconstruction accuracy [18-21]. SENext [22] a Squeeze-and-Excitation Next approach for SISR. Network architecture incorporates squeeze-and-excitation blocks (SEB) to reduce computational costs and dynamically recalibrate channel-wise feature mappings. Local, sub-local, and global skip connections are utilized between each SEB to enhance feature reusability and stabilize training convergence. Also, in the pre-processing step, introduce post-upsampling later instead of the conventional hand-designed bicubic upsampling. SENext outperforms all previously proposed methods. To improve SR performance, Kim et al. [23] proposed a popular SISR technique, a very deep network (VDSR) with twenty layers. With technological advancements, SR networks have become increasingly complex and powerful in exploiting intra-view information. While Zhang et al. [24] fuse residual connections with dense connections to introduce the residual dense network (RDN) for fully exploiting hierarchical feature characterization. More recently, SISR has shown further enhanced performance through the development of residual channel attention networks (RCAN) [25], Residual non-local attention networks (RNAN) [26], and Second-order attention networks (SAN) [27]. Muhammad et al. [28] introduce a novel architecture inspired by ResNet and Xception networks for super-resolution (SR) tasks. The approach significantly reduces network parameters and improves processing speed while achieving high-quality HR images with intricate textures and sharp edges. Experimental results proved the superior performance of the proposed technique in terms of accuracy, speed, and visual quality, establishing it as a state-of-the-art SR method. The SISR technique based on sparse representation, on the other hand, employs sparse coding theory to encode LR and HR image blocks as dictionaries. Then a machine learning model is used to determine the mapping connection between these dictionaries. This connection creates an HR image from a single LR frame.

Developing a strong and efficient learning model and choosing an adequate training dataset are crucial elements that influence how effectively learning-based super-resolution approaches function.

### B. Stereo Image Super-Resolution

Stereo image SR, cross-view information is used from stereo images in SISR to elevate the reconstruction quality. Recently, SISR approaches based on deep learning have demonstrated promising results regarding reconstruction. Yan et al. [29] introduced a domain adaptive stereo SR network that estimates disparities using a pre-trained stereo matching network and utilized cross-view information by warping views to the other side. Xu et al. [30] proposed bilateral grid processing into CNNs to introduce a bilateral stereo super-resolution network (BSSRnet) for stereo image SR. In contrast, Chu et al. [31] introduced NAFSSR new and unique CNN-based NAFNet and a Stereo Cross Attention Module (SCAM) block for parallax fusion. However, existing methods often treat cross- and intra-image features as independent processes, which shrinks their ability to fully exploit image features. Additionally, the match between the baseline networks and parallax fusion blocks is often not considered, leading to sub-optimal results.

### C. Video Super-Resolution (VSR)

Liao et al. [32] presented a Convolutional Neural Network (CNN) for video Super-Resolution (SR), marking a significant breakthrough in the field of SR. Their approach involved using motion compensation to produce a collection of SR drafts. Subsequently, they use a CNN architecture to reconstruct HR frames based on the ensemble of drafts. This methodology represents a novel contribution to the advancement of video SR techniques. Tao et al. [33] devised a novel method by combining an encoder-decoder network with LSTM (Long Short-Term

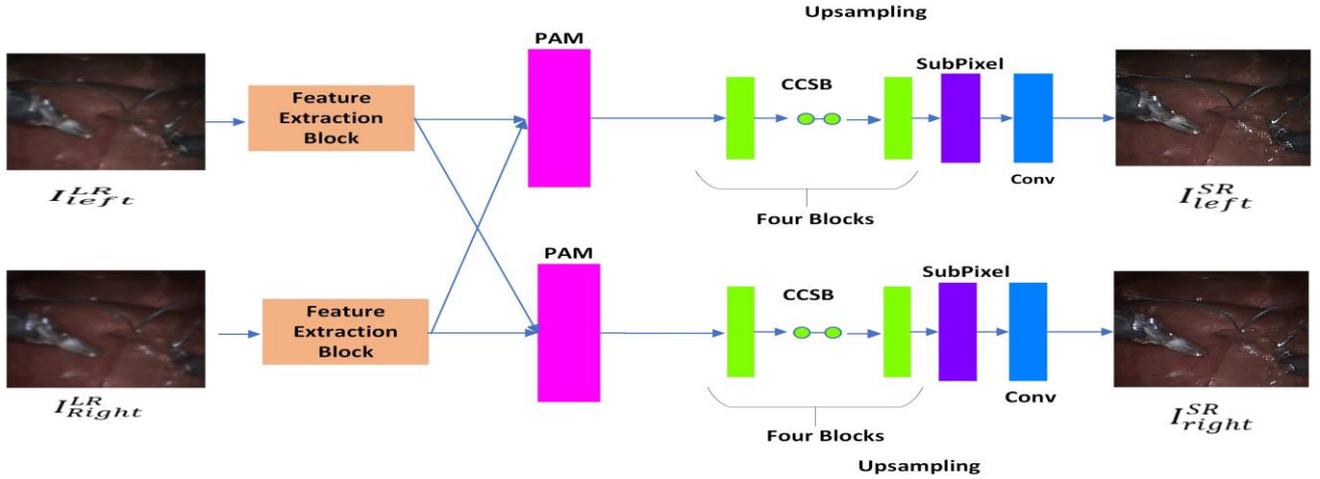

Fig. 1. Overview of Combined Channel and Spatial Attention based Stereo Endoscopic Image Super-Resolution Architecture

Memory) to effectively leverage temporal correspondence in video Super-Resolution (SR). By combining these components, the architecture enabled the extraction of valuable temporal context.

## III. PROPOSED METHOD

Here we'll detail our combined channel and spatial attention block, discuss PAM for super-resolution, and present our loss function. Fig.1 presents an overview of the proposed Combined Channel and Spatial Attention-based Stereo Endoscopic Image Super-Resolution Architecture.

### A. Combined Channel and Spatial Attention Block

The combined channel and spatial attention block (CCSB) are made up of channel attention block (CAB) and spatial channel attention block (SAB) block, as shown in Fig.3. Channel attention helps assign significance to different feature maps, and spatial attention determines important regions within each feature map. Average pooling and max pooling operations are performed simultaneously to combine and compress features, resulting in max pooled and average pooled features. Both max pooled, and average pooled features are fed two dense layers for training. We have adopted a reduction parameter to reduce parameters, and $n_{channels}/r \times 1 \times 1 \times 1$ is set as the activation size. Finally, sigmoid is applied, and we get channel attention ($F_{cal}$, $F_{car}$). To focus on important regions in feature maps, we explore spatial attention. $\otimes$ denotes element-wise multiplication. Refined features extracted from channel attention are fed into max and global average pooling to create a 3-dimensional feature map. The output of both pooling operations is concatenated, and 3-dimensional convolution is applied to create a three-dimensional spatial attention map. The kernel size for convolution is $3 \times 3 \times 3$. Sigmoid is applied to obtain optimized features ($F'_{left}$, $F'_{right}$). LR images are passed into CAB then extracted features from CAB are passed into SAB.

$$F_{cal}=f_{cafe}(I_{left}), F_{car}=f_{cafe}(I_{right}) \qquad (1)$$

$$F_{left}=f_{safe}(F_{cal}), F_{right}=f_{safe}(F_{car}) \qquad (2)$$

$F_{cal}$ is features extracted from the channel attention block when the left view image is processed, while $F_{car}$ is the output of the channel attention block when the right view image is passed through the channel attention block. $F_{left}$ is the output feature when $F_{cal}$ is passed through the spatial attention block. Similarly, $F_{right}$ is an output feature when $F_{car}$ is fed into a spatial attention block. $f_{cafe}(.)$ and $f_{safe}(.)$ are feature extraction module.

### B. Feature Extraction Block

Inspired by Wang et al. [34], a set of three enlarged convolutions (dilation rates of 1, 4, 8) are combined to create an Atrous Spatial Pyramid Pooling (ASPP) group, which is then connected in a residual manner to produce a cascade of three ASPP groups. The ASPP block contribution is an increased receptive field and a wider variety of convolutions. This results in a combination of convolutions with different open regions and dilation rates, enhancing the diversity of the network. The residual ASPP module as shown in Fig. 2 can learn highly discriminative features, significantly enhancing the overall SR performance. The ASPP module is the makeup of a sequence of a residual ASPP block and a residual block, alternately cascaded. The features from CCSB are passed through a residual ASPP block to produce features at multiple scales. Subsequently, these features are fed to a residual block to perform feature fusion. Two such blocks are used to produce the final features. The combined types of blocks allow models to extract and combine multi-scale features, leading to improved performance in super-resolution tasks. CCSB and three ASPP constitute one feature extraction module as shown in Fig. 2.

$$M_{R \to L}, F_{PAM,left} = f_{PAM}(F_{left}, F_{right}) \qquad (3)$$

$$M_{L \to R}, F_{PAM,right} = f_{PAM}(F_{right}, F_{left}) \qquad (4)$$

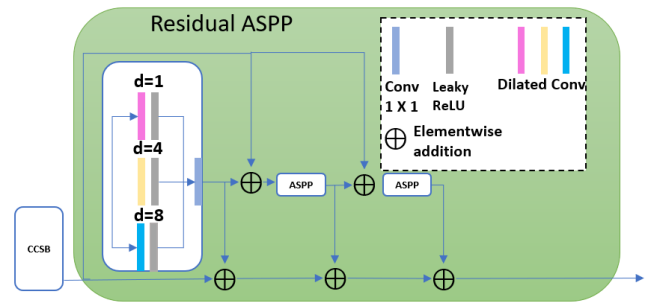

Fig. 2. Residual ASPP

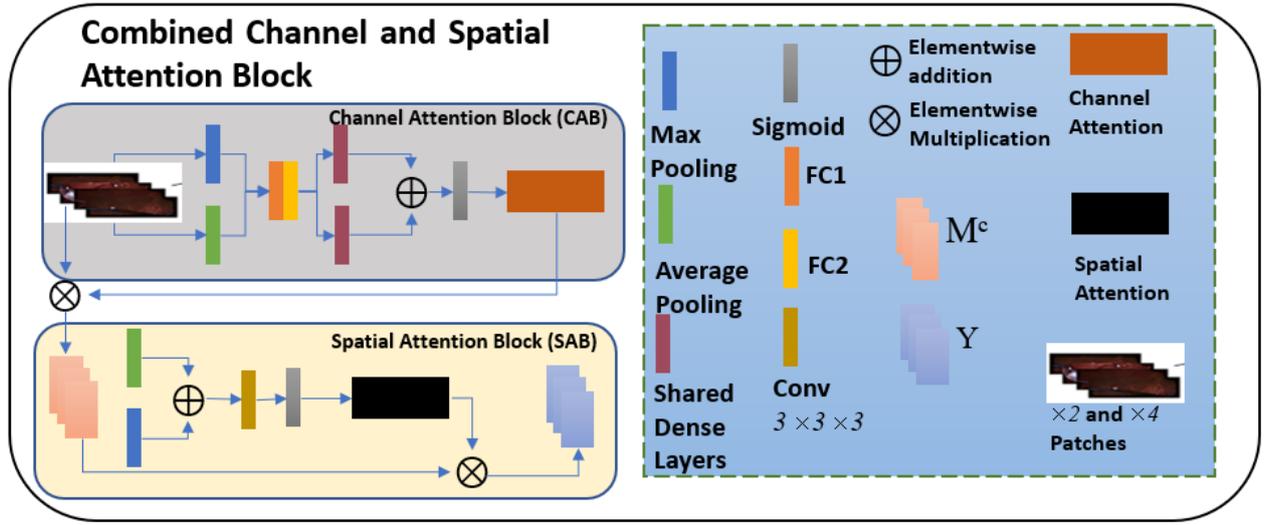

Fig. 3. Combined Channel and Spatial Attention Block

## C. Feature Interaction Module

After feature extraction, the extracted features $F_{left}$ and $F_{right}$ are passed through the feature interaction module to provide the interaction of cross-view information. We follow the concept of [19, 34], to use a PAM block to reduce the various factors of stereo cameras, such as focal length, baseline, and resolutions.

$$I^{LR}_{Left} = M_{R \to L} \circledast I^{LR}_{Right} \quad (5)$$

$$I^{LR}_{right} = M_{L \to R} \circledast I^{LR}_{left} \quad (6)$$

PAM as shown in Fig. 4 is used to obtain the parallax-attention map $M_{R \to L}$ using input features. The attention map from right-to-left $M_{R \to L}$ is transposed to get $M_{L \to R}$, enabling the feature interaction module to communicate information from both simultaneous views.

$$\begin{cases} M_{L \to R \to L} = M_{R \to L} \circledast M_{L \to R} \\ M_{R \to L \to R} = M_{L \to R} \circledast M_{R \to L} \end{cases} \quad (7)$$

After that, we generate valid masks ($V_L$, $V_R$) and concatenate them with attention maps ($M_{R \to L}$, $M_{L \to R}$) and input features ($F_{left}$, $F_{right}$) to create the output features. This approach makes effective interaction of information from both views while accounting for differences in disparities that may arise from variations in camera parameters such as baselines, focal lengths, and resolutions.

$$V_{L \to R}(i,j) = \begin{cases} 1, & if \ \sum_{k \in [1,W]} M_{L \to R}(i,k,j) > \tau \\ 0, & Otherwise \end{cases} \quad (8)$$

## D. Upsampling

The upsampling module we have designed comprises four CCSBs, multiple convolutional layers, and a pixel shuffle layer.

$$I^{SR}_{right} = f_{upsampling}(F_{PAM,right}) \quad (9)$$

$$I^{SR}_{left} = f_{upsampling}(F_{PAM,left}) \quad (10)$$

Where $f_{upsampling}(.)$ is upsampling function.

## E. Loss Function

The loss function consists of three main parts for a stereo-matching model: the SR loss ($L_{SR}$), the parallax-attention loss ($L_{PAM}$), and the stereo consistency loss ($L_{stereo}$).

$$Loss = L_{SR} + L_{PAM} + L_{stereo} \quad (11)$$

The SR loss calculated the similarity between the predicted and high-resolution ground-truth stereo images. In contrast, the parallax-attention loss encouraged the model to emphasize the most salient features of the scene. The stereo consistency

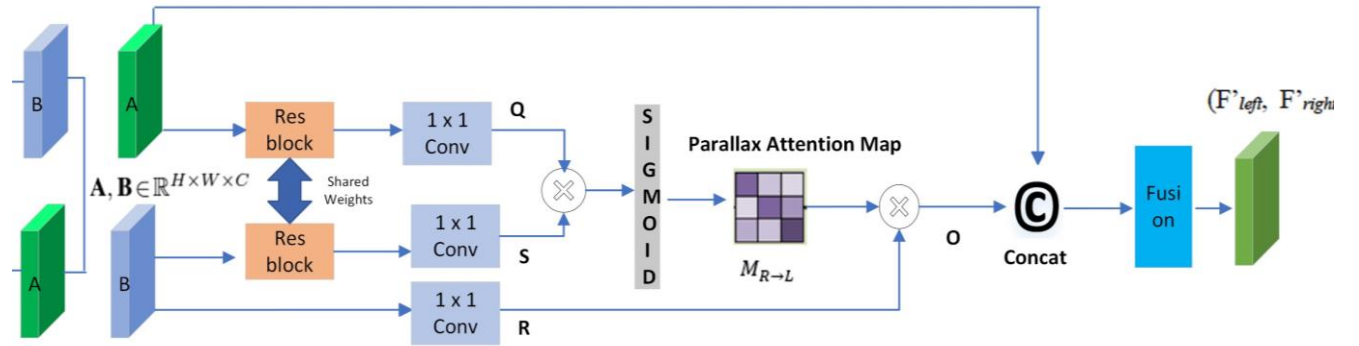

Fig. 4. An overview of Parallax Attention Module (PAM)

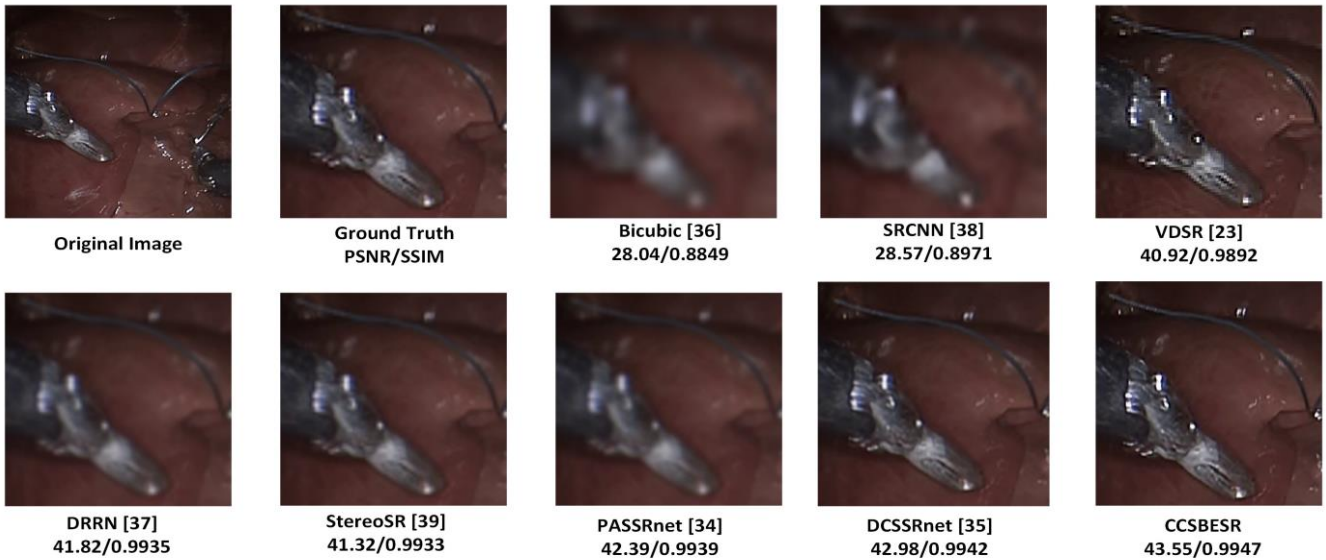

Fig. 5. Perceptual results of different models, corresponding HR image and ground truth image with enlargement scale factor ×2

loss ensures that the predicted depth maps are consistent with the stereo images.

## IV. EXPERIMENTAL RESULTS

### A. Dataset

The da Vinci dataset provided by Zhang et al. [35] available at https://github.com/hgfe/DCSSR, comprises 6300 pairs of endoscopic images, with a resolution of 512 × 512 pixels. These videos are recorded using cameras of da Vinci's surgical system, specifically during surgical operations on the ovine liver. These image pairs were extracted from three different laparoscopic videos with 4560, 870, and 870 image pairs. Scale 2 and scale 4 images are obtained by downsampling original HR images similar to presented in [35]

### B. Training Details

We implemented this model in Pytorch 2.0 and trained using Nvidia 1080Ti GPU. Xavier initializer is incorporated to initialize convolution filters. Adam optimizer is included, and the initial learning rate is set to $3\times10^{-4}$. The model is trained for 40 epochs each with batch size 8 for scale 2 and batch size 2 for scale 4.

### C. Quantitative evaluation

PSNR is a commonly and broadly used objective metric in image SR, indicating the similarity level between the HR and SR images. SSIM is a widely used perceptual metric for measuring image similarity. We compare our model to other super-resolution techniques with similar structures. The average scores tested are presented in Table 1 as the quantitative results. Our algorithm attains the best performance as compared to existing state-of-the-art methods.

### D. Qualitative evaluation

The comparison of SR performance among various approaches is presented in Fig. 5. We compared our proposed model with state-of-the-art techniques. Bicubic [36], DRRN [37] and SRCNN [38] failed to produce better-quality images. While StereoSR [39], PASSRnet [34] and DCSSRnet [35] have shown comparatively better results, there were still some inaccuracies due to artifacts. Our model CCSBESR accurately estimates disparities and handles inaccuracies without introducing artifacts. Our model has improved the image details, e.g., texture and edges.

TABLE I. QUANTITATIVE COMPARISON USING PSNR/SSIM ON DA VINCI DATASET WITH ENLARGEMENT FACTOR ×2 AND ×4

| Methods | da Vinci Dataset | | | |
|---|---|---|---|---|
| | ×2 | | ×4 | |
| | PSNR | SSIM | PSNR | SSIM |
| Bicubic [36] | 37.56 | 0.9912 | 30.07 | 0.9581 |
| SRCNN [38] | 41.38 | 0.9948 | 32.29 | 0.9694 |
| VDSR [23] | 42.02 | 0.9953 | 33.39 | 0.9751 |
| DRRN [37] | 41.93 | 0.9952 | 33.31 | 0.9750 |
| StereoSR [39] | 41.94 | 0.9952 | 33.18 | 0.9740 |
| PASSRnet [34] | 42.03 | 0.9955 | 33.69 | 0.9762 |
| DCSSRnet [35] | 42.18 | 0.9956 | 33.86 | 0.9770 |
| CCSBESR(Our) | 44.30 | 0.9959 | 35.15 | 0.9778 |

## V. CONCLUSION

Our proposed method has incorporated a masking technique that helps test features' luminance sensitivity. The enhanced depth perception provided by CCSBESR will allow medical doctors and other professionals to understand complex anatomical structures, leading to more accurate diagnoses and improving surgical outcomes.

This paper proposes a feature extraction module comprising a Channel attention block, spatial channel attention block, and three Residual ASPP blocks. Combining these two blocks allows extracting refined features for image super-resolution. These extracted features are then fed into the PAM block for feature interaction and at the end, up-sampling is applied to extract Stereo super-resolute images. Experimental results have shown promising results in PSNR/SSIM and outperformed other current methods.

ACKNOWLEDGMENT

Multimedia Data Analytics and Processing Research Unit, Department of Electrical Engineering, Chulalongkorn University, Bangkok 10330, Thailand; This Research is funded by Thailand Science research and Innovation Fund Chulalongkorn University (IND66210019); The NSRF via the Program Management Unit for Human Resources & Institutional Development, Research and Innovation [grant number B04G640053]; Graduate Scholarship Programme for ASEAN or Non – ASEAN Countries.